\documentclass[prl,showpacs,twocolumn]{revtex4}

\usepackage{mathrsfs}
\usepackage{stmaryrd}
\usepackage{times}
\usepackage{amsmath}
\usepackage{amsfonts}
\usepackage{amssymb}
\usepackage{graphicx}
\usepackage{bm}
\begin{document}

\title{Interference Effect of Majorana Fermions in a Spin-orbit Coupled Superconducting Wire}
\author{Jie Liu$^{1}$}
\author{Juntao Song$^{2}$}
\author{Qing-Feng Sun$^{3,4}$}
\author{X. C. Xie$^{3,4}$}

\affiliation{$^{1}$ Department of Applied Physics, School of Science, Xi¡¯an Jiaotong University, Xi¡¯an 710049, China}
\affiliation{$^{2}$Department of Physics and Hebei Advanced Thin Film Laboratory,
Hebei Normal University, Shijiazhuang 050024, China}
\affiliation{$^{3}$International Center for Quantum Materials, School of Physics, Peking University, Beijing 100871, China}
\affiliation{$^{4}$Collaborative Innovation Center of Quantum Matter, Beijing 100871, China}

\begin{abstract}
Two majorana Fermions (MFs) localized at the two ends of the topological superconducting wire
can interfere with each other and form the well known $4\pi$ Josephson current.
We reveal that the density of states (Dos) for the electron part and the hole part
also follow a parity correlated $4\pi$ period oscillation, while the Dos displays a $2\pi$ period oscillation when
two trivial states interfere with each other. Thus, the period of Dos oscillation can be used to distinguish the MFs from the
trivial localized states. Interestingly,
such phenomena can be directly observed in a short superconducting wire controlled by the gate voltage.
This largely simplifies the experimental setup. We suggest
that the interference effects can be detected through two STM leads or two norm leads.

\end{abstract}


\maketitle

{\bf \emph{Introduction}} --- Following the suggestion of Kitaev that MFs can appear as quasi-particle states at the ends
of 1 dimensional (1D) p-wave superconductor[\onlinecite{kitaev}],
how to realize MFs in laboratory becomes a booming focus in condensed matter physics[\onlinecite{nayak}].
A number of proposals have been suggested to fabricate and detect the MFs in effective 1D p-wave superconductor system[\onlinecite{ fujimoto, sau, sato, alicea2, lut, oreg,potter}].
Among these proposals, a semiconductor wire with Rashba spin-orbit coupling and proximity induced superconductivity
is deemed as the most promising choice[\onlinecite{sau}].  Indeed, the semiconductor superconducting nanowire
has been manufactured rapidly to response the prediction of theory[\onlinecite{kou, deng, das1}]. Next to the semiconductor system, the second
topological superconducting system realized in experiment is related to ferromagnetic atomic chains being put on a trivial  superconductor[\onlinecite{perge}].
Both systems detect MFs through tunneling experiment.
It is believed that MFs can cause zero-bias conductance peak (ZBP) in conductance spectrum[\onlinecite{law, wimmer}]
and indeed the experiment has observed the signature of ZBP.
However, ZBPs itself cannot imply the MFs conclusively.
Other sources can cause the similar phenomena, such as the disorder induced trivial states or weak-antilocalization[\onlinecite{jie, altland, piku, kells,tewari}].
To unambiguously determine MFs, further efforts are greatly needed.

Besides ZBPs, another significant feature of MFs is $4\pi$ Josephson current.
When two topological superconducting wire put together to form a topological Josephson junction (Top-JJ),
the super current is $4\pi$ periodic if there are MFs existing at the ends of both wires.
This is different from the trivial case without MFs. In trivial case only Cooper pairs can tunnel, the periodicity
is $2\pi$. MFs enable the tunneling of single electron in a Top-JJ. In this situation, the periodicity is doubled.
Since the  $4\pi$ Josephson effect is an unique transport property of MFs, many groups devote to realize it.
However, Combining two topological superconducting wire together to form a Top-JJ is certainly
a huge challenge to experimentalists. To date, there is no plausible $4\pi$ signature observed. Kouwenhoven's group
have fabricated such junction but failed to observe the $4\pi$ period[\onlinecite{kou1}].
 Several groups have resorted to superconductor-topological insulate-superconductor junction which can also display
 the $4\pi$ Josephson current. However, they only report
  the signature of edge state's information[\onlinecite{yaco,kou2,lu}].
  To realize the $4\pi$ Josephson current, some physical measurement beyond the super current
  may be needed.


\begin{figure}
\centering
\includegraphics[width=3.25in]{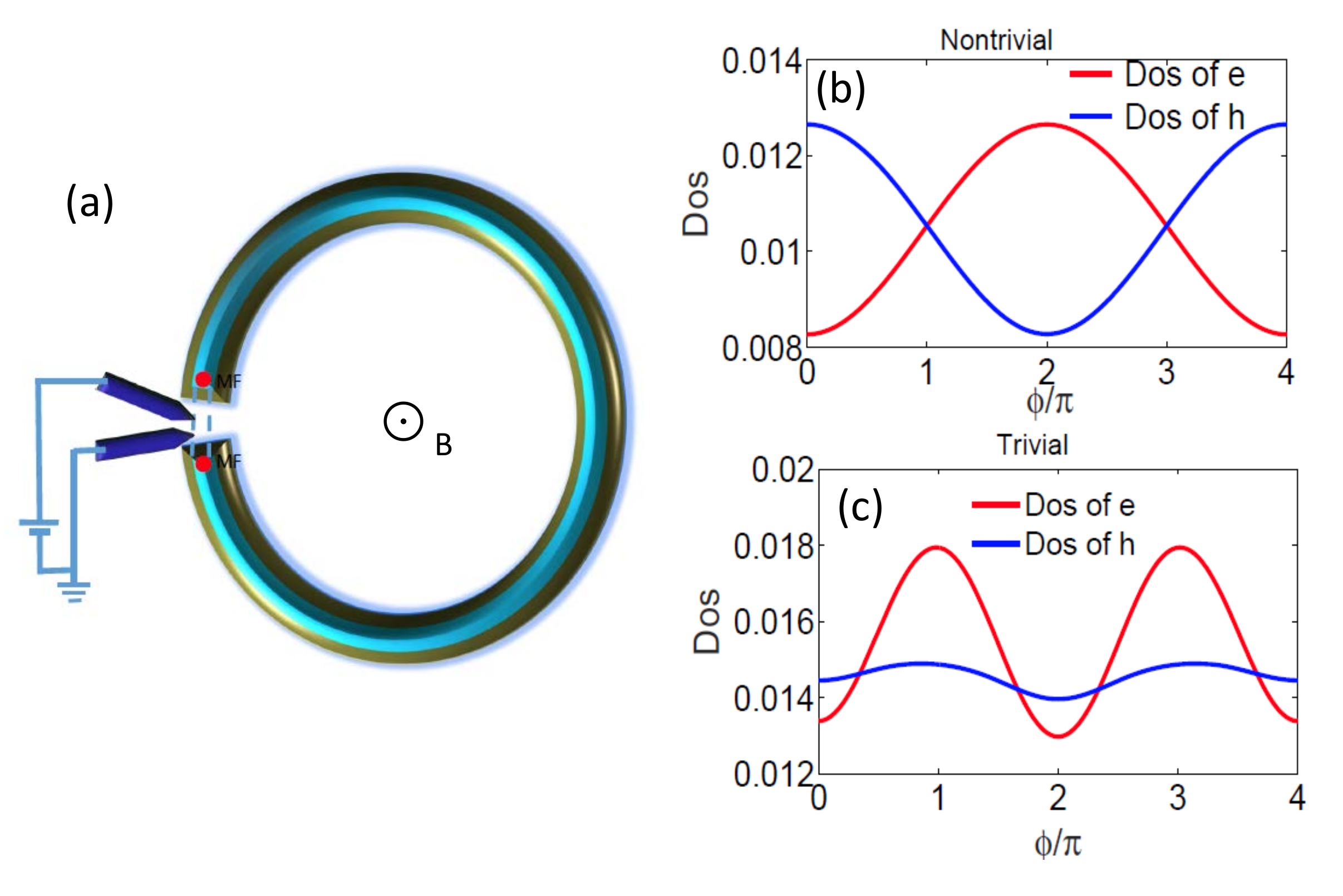}
\caption{ (a) A schematic setup of the experiment, two STM leads or normal leads are coupled to the ends of a superconductor ring which supports MFs. (b) Density of states of coupled Majorana fermion in a superconducting semiconductor ring system. Both electrons and holes show a 4 $\pi$ oscillation though the total density is unchanged with varying of flux. (c) In the trivial region, two trivial states which localized at the two ends of the wire also show the oscillation behavior with varying of the flux. However, the period is an ordinary $2\pi$.}\label{f1}
\end{figure}

In this Letter, we study the Top-JJ in a ring heterostructure wire as shown in Fig. 1(a).
Unlike previous studies, here we focus on the Dos for
the electron part and the hole part. An essential property of MFs is that its wave function of the electron part must be conjugate to its wave function of the hole part, namely the self-hermitian property of MFs. Focusing on the Dos of the electron part and the hole part can directly manifestly the self-hermitian property of MF, this is the basic starting point in this Letter. It is very interesting that the Dos for the electron part and the hole part also display the $4\pi$ oscillation.
        This is because the origin of $4\pi$ Josephson current is caused by the interference of two MFs localized at the two ends of the wire.
        Thus both Dos for the electron part and the hole part show the information of interference. This gives
        a way to identify the $4\pi$ periodicity through the local Dos. We suggest this information can be read through
        two STM leads. Since the $4\pi$ Josephson effects is caused by the interference of two MFs, the similar information can be observed in a short superconducting wire system. There two MFs are also localized at the two ends of the wire, respectively. When the length of the wire is compared with the coherence length of the MFs, two MFs can interfere with each other through the superconducting wire. The more interesting thing is that the phase difference can be adjusted through the gate voltage, which certainly simplifies the requirement of an experiment. A normal lead-superconductor-normal lead (N-S-N) heterostructure can read the interference effect through the electron transmission.

{\bf \emph{Model}} --- We construct a quasi 1D TS wire and study the interference effect of MFs. The TS wire used is a quasi-one dimensional s-wave superconductor with Rashba spin-orbit coupling. Following  Ref.[\onlinecite{potter, jie}], the tight-binding model is:
\begin{eqnarray}\label{model}
 H_{q1D}& =& \sum\nolimits_{\mathbf{R},\mathbf{d},\alpha } { - t(\psi _{\mathbf{R} + \mathbf{d},\alpha }^\dag  \psi _{R,\alpha }  + h.c.) - \mu \psi _{\mathbf{R},\alpha }^\dag  \psi _{\mathbf{R},\alpha } } \nonumber \\
&+& \sum\nolimits_{\mathbf{R},\mathbf{d},\alpha ,\beta } { - i{U _R} \psi _{\mathbf{R} + \mathbf{d},\alpha }^\dag  \hat z \cdot (\vec{\sigma}  \times \mathbf{d})_{\alpha \beta }   \psi _{\mathbf{R},\beta } } \nonumber  \\
 &+& \sum\nolimits_{\mathbf{R},\alpha ,\beta } { \psi _{\mathbf{R}, \alpha }^\dag [(V_x \sigma_x)_{\alpha \beta} +V_{\text{imp}}(\mathbf{R})\delta_{\alpha \beta}]\psi _{\mathbf{R},\beta } } \nonumber \\
& +& \sum\nolimits_{\mathbf{R},\alpha} \Delta \psi _{\mathbf{R},\alpha }^{\dagger} \psi _{\mathbf{R},-\alpha }^{\dagger}+h.c. \\
H_{end} & = & \sum\nolimits_{\mathbf{i_y},\alpha }(t_c e^{-i\phi/2} \psi _{N_x,\mathbf{i_y},\alpha}^{\dagger} \psi _{1,\mathbf{i_y},\alpha}+h.c.).
 \end{eqnarray}
Here, $\mathbf{R}$ denotes the lattice sites, $\mathbf{d}$ denotes the two unit vectors $\mathbf{d_{x}}$, $\mathbf{d_y}$ which connect the nearest neighbor sites in the $x$ and $y$ directions respectively\cite{note1}. $\alpha, \beta$ are the spin indexes. $t$ is the hopping amplitude, $\mu$ is the chemical potential, $U_{R}$ is the Rashba coupling strength, $V_{x}$ is the Zeeman energy caused by a magnetic field along the wire direction. $\Delta$ is the superconducting pairing amplitude and $V_{\text{imp}}(\mathbf{R})$ is  the on-site random impurity.
$H_{end}$ means the coupling between the two ends of the wire (here we bend the wire to form a ring)
 and $\phi$ is the flux in the ring. To match the experiment in Ref.[\onlinecite{jie}], the parameters in the tight-binding model are chosen as follows:
   $\Delta=250 \mu \text{e}V$, $t=25\Delta$, $U_{R}=2\Delta$, and the superconductor coherence length $\xi=t/\Delta a= 500 nm$. In addition, we set $V_x=2\Delta$ such that the superconducting wire can support the MF end states by tuning the chemical potential.


{\bf \emph{$4\pi$ oscillation of the density of states}}---It is well known that MFs obey $4\pi$ fractional Josephson effect.
Here we show that the Dos of electron (hole) part of Andreev bound states formed by two MFs also oscillate with  $4\pi$ period.

The  $4\pi$ period is directly related to the self-hermitian and fractional nature of MFs. Self-hermitian requires that the wave function of electron part must be the conjugate of the wave function of hole part .
Thus,  a general wave function of MFs should be: $
 \psi_n =  \left(e^{i\phi_n/2},  e^{-i\phi_n/2} \right)^{T} e^{-x/\xi}.$
 In a Top-JJ as shown in Fig. 1 (a), two MFs will couple to each other to form an Andreev bound states, the excited wave function should be:
 \begin{eqnarray}
 \psi_{\pm} = \psi_1-i(-1)^{N_v} \psi_2 =  \left(\begin{array}{ll} 1\mp ie^{i\phi/2} \\ 1\mp  ie^{-i\phi/2} \end{array}\right) = \left(\begin{array}{ll} u_{\pm}\\ \upsilon_{\pm } \end{array}\right) .
\end{eqnarray}
 Here, $\phi=\phi_2-\phi_1$, $N_v=0,1$ corresponding to the odd and even states of the system $E(\phi) \varpropto (-1)^{N_v}  sin(\phi/2)$\cite{note2}. We can get the Josephson current
$I_J = \partial E(\phi)/\partial \phi \varpropto (-1)^{N_v}cos(\phi/2)$, which shows a $4\pi$ periodic oscillation. However,  $4\pi$ needs a more stringent condition that requires the parity conservation\cite{law1}£¬namely, the evolution of the states should follows one branch of the spectrum. While it is  particularly susceptible at the degenerate point when the even parity state and odd parity state cross the zero energy at $\phi=2n\pi$.  The states will change from one parity state to another another parity state due to the influence of quasiparticle poisoning, background and thermal effect\cite{Fran,houzet,shu}.  In this case, $4\pi$ will return to the usual $2\pi$.  Thus, besides the technical challenges, parity conservation is also a huge challenge in an experiment.

Fortunately,  both Dos of electrons and holes oscillate with $4\pi$ periodicity with or without parity conservation since they are parity correlated.  From Eq.(3) we can see that Dos for electron is $|u_{\pm}|^2 \varpropto 1\pm sin(\phi/2)$ and the Dos for hole is $|\upsilon_{\pm}|^2 \varpropto 1\mp sin(\phi/2)$. Which means we can distinguish the $4\pi$ information by resorting the Dos of electron (or hole) part along one energy spectrum. We do not need to worry about which way to go as the system evolves, it is free to parity conserving problem.  Our numerical results directly verify this conclusion. We used the tight-binding model as described in Eq. (1). The length of the wire is $N_xa=4\mu m$ and $t_c=0.4t$. Fig. 1(b) shows the information of Dos along $E(\phi) \varpropto sin(\phi/2)$ with setting
 the chemical potential $\mu=-2t$ which lies in the topological region. As we adjust the flux $\phi$, both Dos of electrons and holes
oscillate $4\pi$ period and parity correlated. While the Dos of electrons along $E(\phi) \varpropto -sin(\phi/2)$ is the same as the Dos of hole along $E(\phi) \varpropto sin(\phi/2)$.
 Interestingly,
when two trivial fermion states interfered with each other, the situation
is totally different. In Fig. 1(c), when we set the chemical
potential $\mu=-2t+4$ which lies at the non-topological region, we can see the
period is $2\pi$ both for Dos of electrons and holes.

\begin{figure}
\centering
\includegraphics[width=3.25in]{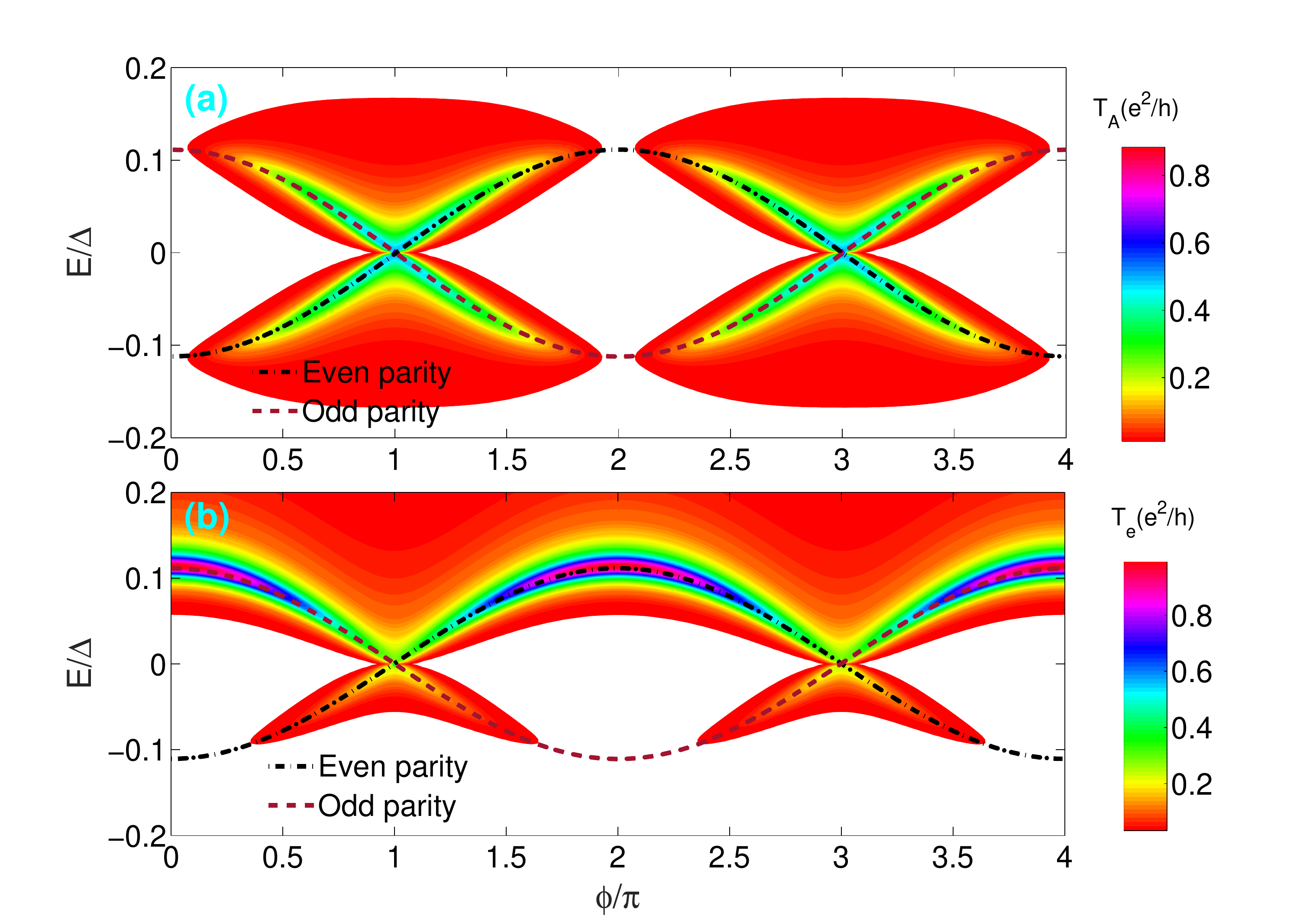}
\caption{Two STM leads localized at the junction of the superconducting ring can read the putative $4\pi$ period through the differencial conductance. (a) Contour plot of Andreev Reflection coefficient of a STM lead versus the flux $\phi$ and incident energy $E$. (b) Contour plot of electron tunneling coefficient $T_{LRe} $ from STM lead L to STM lead R versus the flux $\phi$ and incident energy $E$.  For a single tunneling event, the period is $2\pi$, However, we can distinguish the $4\pi$ period when combined both two tunneling event. The parameters are: $Nx=200a, \mu=-2t, V_x = 2\Delta$.}
\end{figure}

{\bf \emph{Detecting the $4\pi$ oscillation through two STM leads}} ---
In the last section we have shown that the Dos of electrons or holes in topological region is $4\pi$ periodic.
 A nature question arising is how to detect the $4\pi$ periodicity of Dos. Naive method is to put a STM lead (or normal lead) to
 detect the local density. In reality, this does not work. When a STM lead is put in the junction, we do detect a butterfly pattern conductance as we vary the flux $\phi$ in Fig. 2(a) which can be deemed as an unique property of MFs in our previous paper\cite{jie1}.
 However, the peak value of the butterfly for each parity-conserved energy spectrum
  is $2\pi$ periodic instead of $4\pi$ periodic. The reason is that a single STM lead can only read the information of
  local Dos through Andreev reflection(AR).
  The coefficient of AR is $T_A = \frac {\Gamma_e \Gamma_h}  {(\omega -E_M)^2+(\Gamma_e+\Gamma_h)^2} $\cite{jie2}. Here, $\Gamma_e $ is the self-energy of electron part of the leads while $\Gamma_h$ is the self-energy of hole part of the leads, and $E_M$ is the coupling energy of two MFs.
  $\Gamma_e \varpropto |u_{\pm}|^2= 1\pm sin(\phi/2)$ is proportional to the Dos of electron part and $\Gamma_h \varpropto |\upsilon_{\pm}|^2=1\mp sin(\phi/2)$ proportional to the Dos of hole part. Thus,
  AR read the combined Dos of electron and hole part. What's more, we can see that if two MFs are uncoupled to each other , then $|u_{\pm}|^2 = |\upsilon_{\pm}|^2$ and
  $T_A$ will show the well known resonant AR caused by MFs. To detect the local Dos of electron part and hole part, we need additional information beyond AR process.
  Thus, adding another STM lead to detect the electron transmission (ET) or crossed Andreev reflection(CAR) between the two leads is essential\cite{jie2,nilsson}, which can directly manifest the information of Dos. In such process, The tunneling coefficient for ET $T_{e} =  \frac {\Gamma_{Le} \Gamma_{Re}}  {(\omega -E_M)^2+(\Gamma_{Le}+\Gamma_{Lh}+\Gamma_{Re}+\Gamma_{Rh})^2} $.
  Here $\Gamma_{L(R)e} \varpropto |u_{\pm}|^2= 1\pm sin(\phi/2)$ is the  electron part self-energy of STM lead L(R) which is proportional to the local density of states for electron part.
  In Fig 2. (b) we show the contour plot of $T_{e}$ versus flux $\phi$ and incident energy E.
  We can see that the peak value of tunneling coefficient $T_{e}$ is proportional to $(1-sin(\phi/2))^2$. Combine both AR and ET, we can know that the Dos of electron part is maximum at $\phi=2\pi$ and  is minimum at $\phi= 0$ for even parity energy spectrum of two coupled MFs while it is maximum at $\phi= 0$ and is minimum at  $\phi=2\pi$ for odd parity energy spectrum of two coupled MFs. Thus, both branches of Dos show the parity correlated $4\pi$ oscillation.

\begin{figure}
\centering
\includegraphics[width=3.25in]{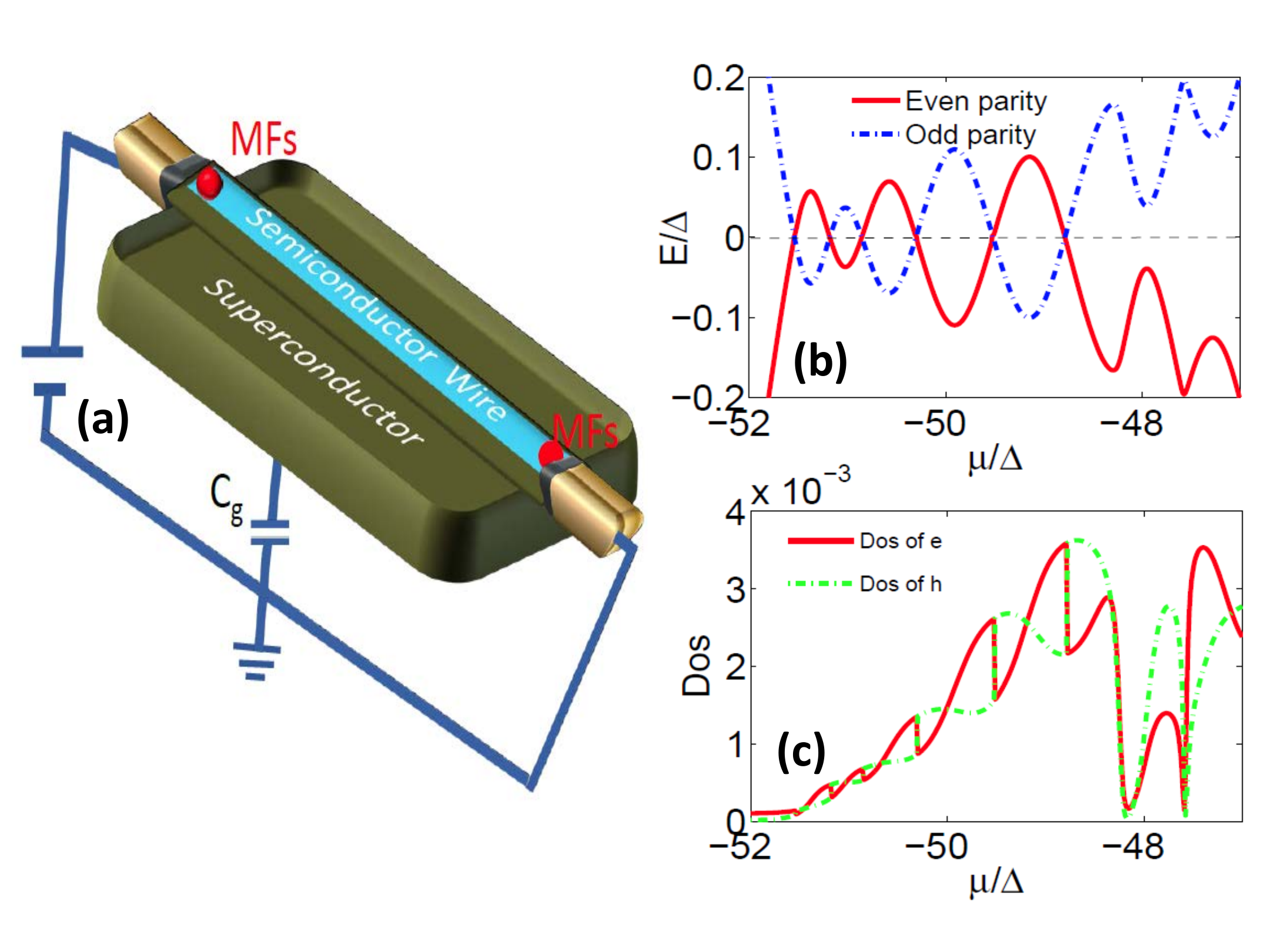}
\caption{A short semiconductor superconducting wire structure can show similar interference effect as the semiconducting superconducting ring. Unlike the superconducting ring system which needs an additional flux to control the interference of Majorana Fermions. The short semiconductor superconducting wire can be easily controlled by gate voltage or chemical potential of the TS. (a) A schematic setup of experiment, two normal leads are coupled to the two ends of a short superconductor wire. (b) The energy spectrum show an oscillation behavior  versus the chemical potential $\mu$.
(c) The Dos of electron part and hole part which localized at $x=1$ versus the chemical potential $\mu$. we set the parameters as: $Nx = 50a, V_x = 2\Delta$.}
 \end{figure}

{\bf \emph{Similar oscillation behavior in a short semiconductor superconducting wire}}---Since the origin of $4\pi$ Josephson current is caused by the interference effect of two MFs.  Another interesting question we want to ask is that whether the interference effect in a short topological superconducting wire is the same as the interference effect in the Top-JJ. In this case the MFs can interfere with each other through the wire.  Actually, there are a number of papers\cite{jie2,nilsson,das} consider about the short wire case. However, most of previous works focus on the non-local transport property of MFs, none of them studies the interference pattern of Dos for electron part and hole part. Only Ref.\cite{das} has focused on the coupling energy of two MFs in a short wire.  They show that the coupling energy oscillate with chemical potential
  $E_M \approx (-1)^{N_v}k_{F}\frac{e^{-L/\xi}} {\xi}sin(k_{F}L)$. Here, $k_{F}\approx \sqrt{\sqrt{V_x^2-\Delta^2}+\mu}$ is the effective fermi wave vector
  which is the function of chemical potential $\mu$ or Zeeman field $V_x$. Thus, the phase difference $k_{F}L$ can be adjusted through the gating voltage or Zeeman field. Thus, we suggest a setup as shown in Fig. 3(a) which consists of a short semiconductor superconducting wire with a gate voltage that controls the chemical potential. Fig. 3(b) shows the energy spectrum varying with the chemical potential, the coupling energy is oscillating  with the chemical potential as Ref.\cite{das} revealed. The key question is whether the parity correlated Dos of electron and hole part is still hold. The answer is yes.
However, there are several differences compared to a long superconductor ring.
 First,  there is a $\pi$ phase shift due to the fact that the wave function has reversed the direction.
 Second,the wave function must propagate a length $L$ to interference with each other. Due to these reasons, the
  Dos of electron part at the ends is $|u_{end}|^2=|1+(-1)^{Nv}e^{(L)/\xi}e^{\pm ik_{F}(L)}|^2=(1+e^{-2L/\xi}+(-1)^{Nv}2e^{-L/\xi}cos(k_{F}L))$.
 In Fig. 3(d), we show the Dos of electron part and hole part at the ends of the wire along spectrum $E_M \propto |sin(k_FL)|$ with increasing $\mu$, which means that the states
change the parity each time when $E_M$ crosses the zero energy. We can see that the interference pattern of Dos of electron or hole part will change correspondingly and it is more obvious due to the $\pi$ phase shift of the interference. The Dos will experience a sudden change with
parity change.

  In the last section we have shown that similar interference effect can happen in a short superconducting wire
and can be controlled by chemical potential or the Zeeman field.  In the following, we discuss how to detect the parity related oscillation behavior of Dos.
 As we have shown, two leads are essential for detecting the whole information of Dos.
Thus we suggest an N-S-N structure for the detection as shown in Fig. 3(a).
Fig. 4(a) shows  the total conductivity can be measured in the right lead. It constitutes two parts: CAR and ET processes. Fig. 4(b) shows the contour plot of ET process $T_e$ versus $\mu$ and $E$.
It clearly show the information of interference.
In one region, the interference effect is constructive and $T_e$ is large, while for another
  region, the interference effect is destructive and the transmission coefficient is very small.
Fig. 4(c) shows the contour plot of CAR process $T_h$ versus $\mu$ and $E$.
However, the peak value of CAR varies little with the
chemical potential and it is hard to see any information about the interference effect.

       In general, the probability of transmission as an electron or a hole is the same due to the self-hermitian of MFs\cite{jie2,nilsson}. While our numerical results show two distinctive results. The reason lies at the novel interference effect of MFs.
       Due to the interference effect of MFs, the Dos of electron part is
 $|u_{end}|^2= 1+e^{-2L/\xi}+(-1)^{Nv}2e^{-L/\xi}cos(k_{F}L)$ while the Dos of hole part is
 $|\upsilon_{end}|^2= 1+e^{-2L/\xi}-(-1)^{Nv}2e^{-L/\xi}cos(k_{F}L)$. As we have shown,
  the probability of CAR process is proportional to the joint density of electron part and hole part,
  namely $T_h \propto |u_{end}\upsilon_{end}|^2 \propto (1+e^{-2L/\xi})^2-e^{-2L/\xi}cos(k_{F}L)^2\approx 1$. Here we consider $exp(-L/\xi)$ as a small quantity and neglect the higher orders.
  Thus, CAR is almost unchanged with the phase difference due to the exponential decay $exp(-L/\xi)$. However, the case for
  ET is different. The coefficient $T_e\propto \Gamma_{Le}\Gamma_{Re} \propto |u_{end}|^2 \propto (1+e^{-L/\xi}cos(k_{F}L))^2 \approx 1+4e^{-L/\xi}cos(k_{F}L)$. Then we can see that the ET process is more sensitive
   to the phase difference and shows an oscillating behavior.

\begin{figure}
\centering
\includegraphics[width=3.25in]{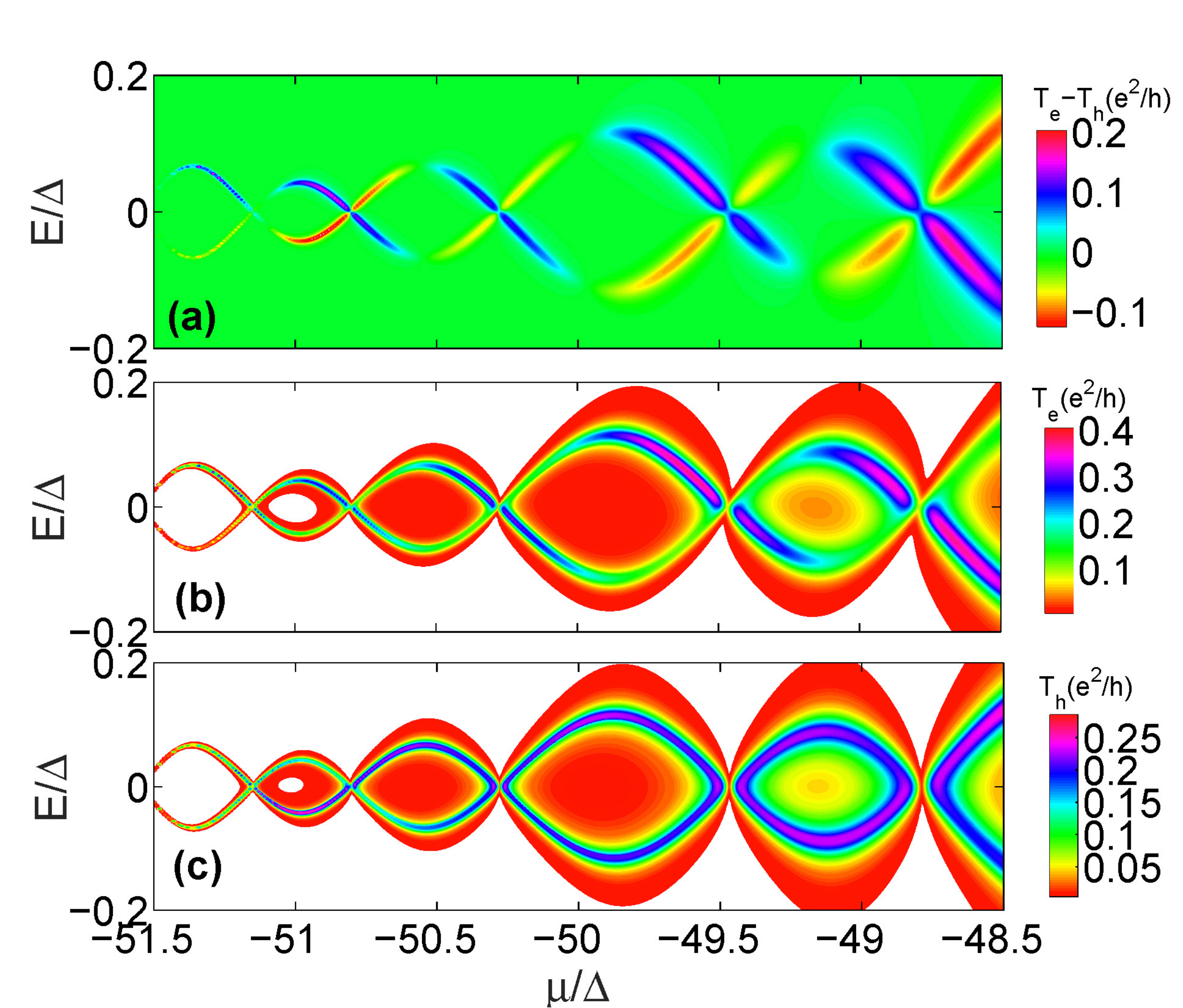}
\caption{ (a) Contour plot of  total differential conductance from the left lead to the right lead with incident energy $E$ and chemical potential $\mu$. (b) Contour plot of electron transmission coefficient $T_{e}$. (c) Contour plot crossed Andreev reflection $T_A$ as a function of  incident energy $E$  and chemical potential $\mu$.  The electron transmission shows a clear oscillation behavior while the CAR is not apparent. This directly demonstrates the self-hermitian property of MFs.  }
\end{figure}

{\bf \emph{Discussion} }--- A general current formulae for a Josephson junction can be described as $j = e(\frac{dn_e}{dt}-\frac{dn_h}{dt})$, here $n_{e(h)}$ means the Dos of electron (hole). If we consider $\phi$ as the time varying parameter, then we can see the Dos of the Andreev bound states is directly related to the well known $4\pi$ Josephson current.
Thus, detecting the $4\pi$ through the density is meaningful. In addition, $4\pi$ needs invariance of the parity conservation while Dos  is free of such restriction. Then it is more simple to detect the $4\pi$ via Dos.

The most surprising thing is that such interference effect can happen in a short wire heterostructure and can be measured through the
ET process. This greatly simplifies the requirement of an experiment.
  Another thing we want to stress is that insensitivity of the CAR process directly manifests the self-hermitian property of MFs. For two trivial states,  the Dos of electron caused by interference effect should be $1+acos(\phi)$ while the Dos of hole should be $1+bcos(\phi)$ (we set $a$ and $b$ as small quantity), in general $T_h\propto 1+(a+b)cos(\phi)$,
 Thus we can still see the interference effect caused by trivial states in CAR process. As for MFs, self-hermitian requires that $a=-b$, then it is hard to see the interference effect in CAR process. Thus,
 the ET process and CAR process in such a short semiconductor superconducting wire
 not only manifest the non-local property of MFs but also manifest the self-hermitian property of MFs.

{\bf \emph{Conclusion} }--- We have shown that the Dos for electrons or holes can directly manifest
 the physics of the $4\pi$ Josephson current, due to the reason that the $4\pi$ Josephson current can be interpreted as the interference of two MFs.
   Therefor, we can identify the $4\pi$ periodicity through the local Dos of electron or hole. The similar physics will appear in a short superconducting wire. We  suggest all the information can be detected by two STM leads or normal leads in the short superconducting wire. In the short wire case, the ET process can show the interference effect clearly while the CAR process is insensitive to such interference effect.
  We show that the insensitivity is protected by the self-hermitian property of MFs.

{\bf \emph{Acknowledgement}}--- We gratefully acknowledge the
support from NSF-China under Grant Nos.11574245(J.L.), 11204065(J.T.S.), 11474085(J.T.S.),11274364(Q.F.S.), 11574007(Q.F.S.), and 11504008(X.C.X.), and NBRP of China under Grand Nos. 2012CB921303 and 2015CB921102. Jie is also support by Postdoc Grant Under Grant No. 2015M580828 and Fundemental Research Founds for central University with Grant No. xjj2015059.

 {99}

\newpage
\section{Supplementary Material}

{\bf \emph{ Formula for calculating the current }} --- In main text we use the recursive Green's function method to calculate the scattering matrix of the model \cite{jie2} where the scattering matrix is related to the Green's functions of the superconducting wire by
\begin{equation}
S_{lk}^{\alpha \beta}=-\delta_{l,k} \delta_{\alpha,\beta }+i[ \Gamma_{l}^{\alpha} ]^{1/2}*G^r*[ \Gamma_{k}^{\beta}]^{1/2}.
\end{equation}
Here, $S_{lk}^{\alpha,\beta}$ is an element of the scattering matrix which denotes the scattering amplitude of a $\beta$ particle from lead $k$ to an $\alpha$ particle in lead $l$, where $l,k=L$ or $R$. L and R denote the STM lead 1 and the STM lead 2 respectively.  The electron ($e$) or hole ($h$) channels are denoted by  $\alpha, \beta, \in \{e,h\}$ . $G^{r}$ is the retarded Green's function of the superconducting wire. $\Gamma_{l}^{\alpha}=i[(\Sigma_{l}^{\alpha})^r-(\Sigma_{l}^{\alpha})^a]$, where $(\Sigma_{l}^{\alpha})^{r(a)}$ is the $\alpha$ particle retarded (advanced) self-energy of lead $i$. The conductivity  of Andreev reflection, crossed Andreev reflection and
electron transmission can be got through the scattering matrix.

{\bf \emph{Effective Hamiltonian and Effective current formula}} --- Both the ring geometry and short wire cases there are two MFs localized at the ends of the wire when the system lie in topological region. The two MFs coupling to each other and form an fermion states. If we add two normal leads to  detect the system, Both two systems can be described as a norm lead-superconductor-normal lead (N-S-N) system. The effective Hamiltonian $H_{eff}= H_{L} + H_{M} + H_{T} $ can be given as follows:
\begin{equation} \label{eff}
\begin{array}{l}
H_{N}  = -iv_{f} {\sum\limits_{\alpha \in {L/R} }} \int_{-\infty}^{+\infty}{\psi_{\alpha}^{\dag}(x)\partial_x \psi_{\alpha}(x)} {dx}, \\
H_M  = i E_M \gamma_{1}\gamma_{2} \\
H_{T}  = \sum\limits_{\alpha }-i [\gamma_1( \tilde{t}_{\alpha,1}\psi_{\alpha}^{\dag}(0) + \tilde{t}_{\alpha,1}\psi_{\alpha}(0)) \\
+ \gamma_2(\tilde{t}_{\alpha,2}e^{-i\phi/2}\psi_{\alpha}^{\dag}(0)+\tilde{t}_{\alpha,2}e^{i\phi/2}\psi_{\alpha}(0) )].
\end{array}
\end{equation}
Here, $H_{N}$ is the Hamiltonian of the left and right normal leads, $\psi_{L/R}$ denotes a fermion operator of the left (right) normal lead. $v_{f}$ is the corresponding Fermi velocity of the leads. $H_{M}$ describes the two coupled Majorana fermions, where $E_{M}$ is the coupling strength between the two MF end states  $\gamma_1$ and $\gamma_2$. The coupling between the leads and the MFs are described by $H_{T}$, where the coupling strengths are denoted by $\tilde{t}_{\alpha,1}$ and $\tilde{t}_{\alpha,2}$ respectively.

\begin{figure}
\centering
\includegraphics[width=3.25in]{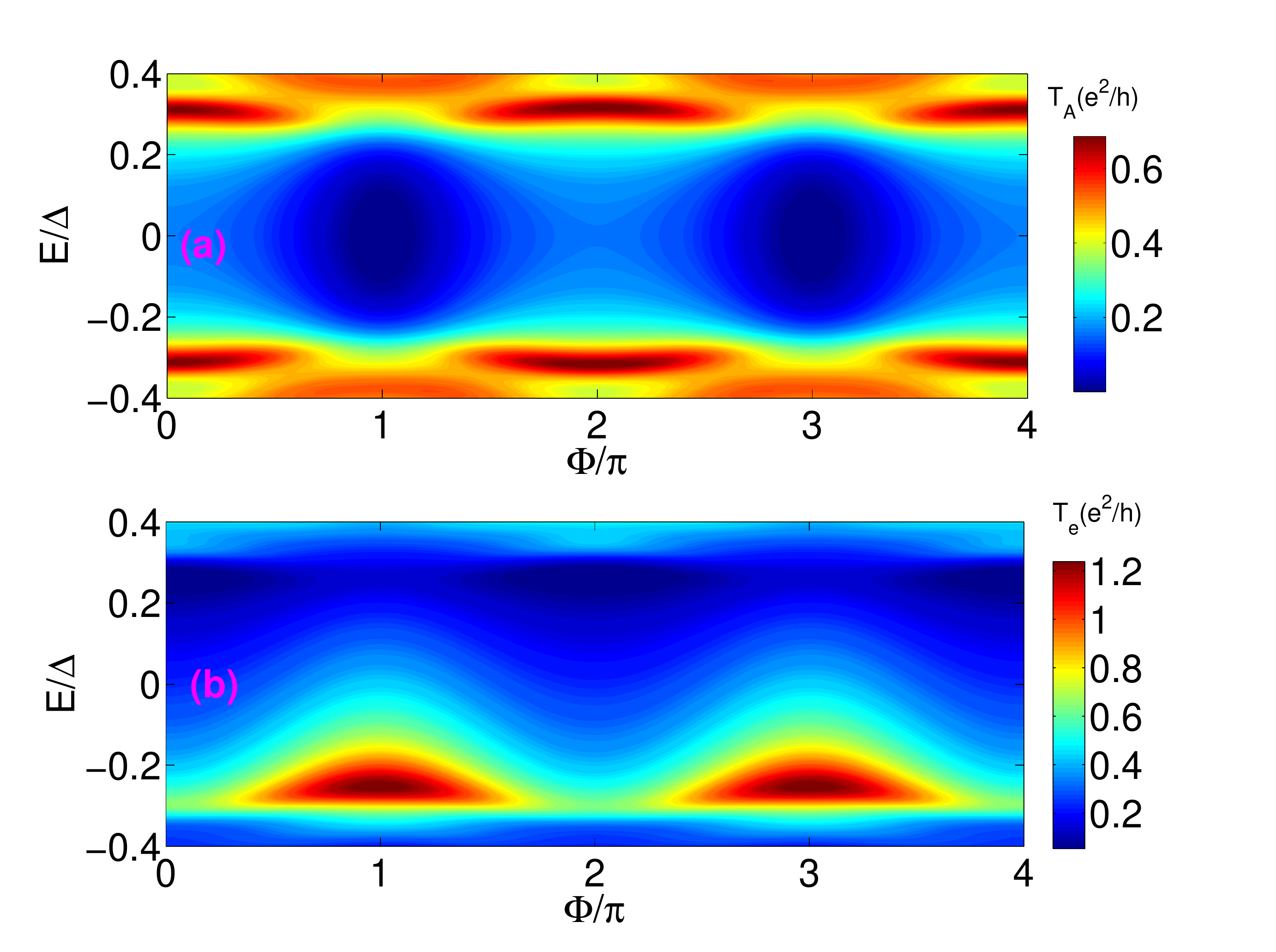}
\caption{The case in the trivial region: (a) Contour plot of Andreev Reflection coefficient of a STM lead versus the flux $\phi$ and incident energy $E$. (b) Contour plot of electron tunneling coefficient $T_{e} $ from STM lead 1 to STM lead 2 versus the flux $\phi$ and incident energy $E$.  Here we can see two maxims peaks at $\Delta\phi = \pi,3\pi$, which shows the Dos of electron part is $2\pi$ periodic. The parameters are: $Nx=200a,  \mu=-2t+4, V_x = 2\Delta$. } \label{Vx}
\end{figure}

To calculate the scattering matrix of the system, we should do a transformation first. Remembering that a single MF is just half of the ordinary fermion states, we can change the MF representation to  fermion representation $\gamma_1 = d+d^{\dagger}, \gamma_2 = i(d-d^{\dagger} )$, then $H_M$ and $H_{T}$ will change to :
\begin{equation} \label{eff1}
\begin{array}{l}
\tilde{H_M}  = E_M d^{\dagger}d \\
\tilde{H_{T}}  = \sum\limits_{\alpha }(  \tilde{t}_{\alpha,e}\psi_{\alpha}^{\dag}(0)d + \tilde{t}_{\alpha,h}\psi_{\alpha}^{\dag}(0)d^{\dagger} +h.c.).
\end{array}
\end{equation}
Here, $\tilde{t}_{\alpha,e}=-i(\tilde{t}_{\alpha,1}+i\tilde{t}_{\alpha,2}e^{-i\phi/2}),\tilde{t}_{\alpha,h}=-i(\tilde{t}_{\alpha,1}+i\tilde{t}_{\alpha,2}e^{i\phi/2})$ (For ring structure it can be write as $\tilde{t}_{\alpha,e}=-i\tilde{t}_{\alpha,1}(1+ie^{-i\phi/2}),\tilde{t}_{\alpha,h}=-i\tilde{t}_{\alpha,1}(1+ie^{i\phi/2})$ while for short wire case $\tilde{t}_{\alpha,e}=-i\tilde{t}_{\alpha,1}(1+ie^{-L/\xi}e^{-i\phi/2}),\tilde{t}_{\alpha,h}=-i\tilde{t}_{\alpha,1}(1+ie^{-L/\xi}e^{i\phi/2})$). Then we can write the scattering matrice in a model-independent form,
\begin{equation} \label{eff2}
S(E) = 1-2\pi i W^{\dagger}(E-\tilde{H_M}+i\pi WW^{\dagger})^{-1}W,
\end{equation}
with $W$ the matrix that describes the coupling of the scattering to the leads:
\[W = \left( {\begin{array}{*{20}{c}}
{{\tilde{t}_{L,e}}} & {{\tilde{t}_{R,e}}} & {{\tilde{t}_{L,h}}} & {{\tilde{t}_{R,h}}}\\
-\tilde{t}'_{L,h} & -\tilde{t}'_{R,h} & -\tilde{t}'_{R,e} & -\tilde{t}'_{R,e}\end{array}}
 \right)\].

In general, we can write the approximation form as : $S_{lk}^{\alpha \beta}=-\delta_{l,k}\delta_{\alpha,\beta}+i\sqrt{\Gamma_{l,\alpha}\Gamma_{k,\beta}}/(E-E_M+i\Gamma)$. Here, $\Gamma_{l,\alpha}$ is the self-Energy of $\alpha$ part of the lead $l$, which is renormalized by the local Dos of the two coupled MFs and proportional  to the local Dos of the two coupled MFs' $\alpha$ part. Thus, from scattering matrix we can read the information of local Dos. However, just a single tunneling process can't give the whole information, we need more tunneling process to read the information. Two leads are necessary in such case.

\begin{figure}
\centering
\includegraphics[width=3.25in]{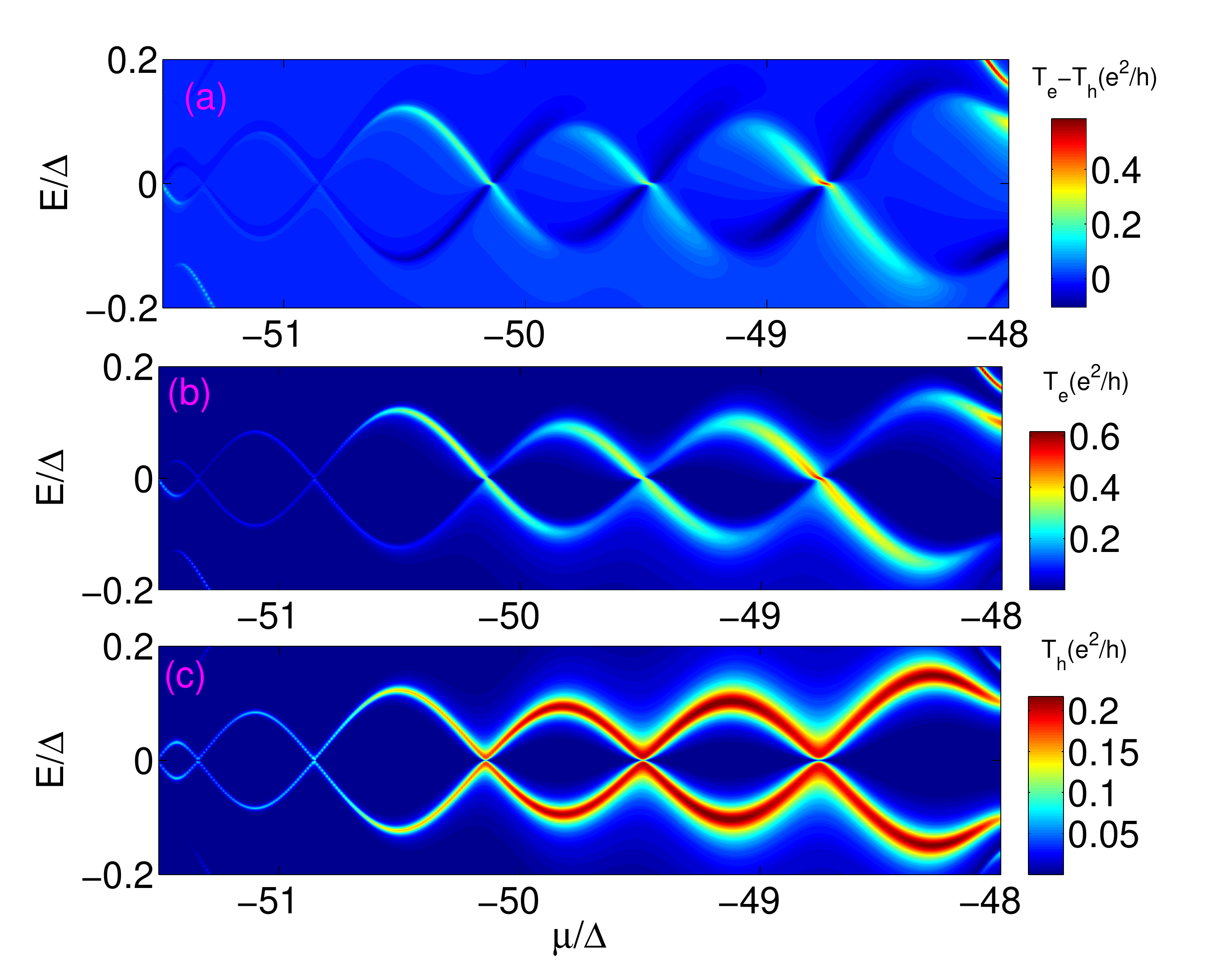}
\caption{ The case with the influence of  disorder, we can see that the system is protected by topology and we can still see the information of interference. The disorder strength is $V_{imp}=4\Delta$ with disorder equally distributed at range $[-V_{imp}/2,V_{imp}/2]$. (a) Contour plot of  total differential conductance from left lead to right lead with incident energy $E$ and chemical potential $\mu$. The total conductance can be described by two parts: $T_e$ contributed by electron transmission and $T_h$ contributed by CAR. (b) Contour plot of electron transmission coefficient $T_{e}$. In this case the whole information uninfluenced by disorder, we can watch the interference effect through the ET process well. (c) Contour plot crossed Andreev reflection $T_h$ as a function of  incident energy $E$  and chemical potential $\mu$.  } \label{multi}
\end{figure}

{\bf \emph{STMs information in trivial region in the ring geometry}} --- In main text we show that
two STM leads can read the information of $4\pi$ in nontrivial region. Here we show the tunneling results in the trivial case with $\mu=-2t+4$. Supplemental Fig. 1(a) show the contour plot of Andreev Reflection coefficient of a STM lead versus the flux $\phi$ and incident energy $E$ while (b) show the contour plot of electron tunneling coefficient $T_{e} $. We can see that it is totally different from the non-trivial case. $T_{e}$ do not cross the zero points and show two peaks at the positional $\pi$ and $3\pi$, which means it is $2\pi$ periodic in trivial region.

{\bf \emph{The case with the influence of disorder}} --- Disorder is unavoidable in real system, it's necessary to consider the influence of disorder. Here  we consider the influence of disorder in short wire case. As demonstrated by our numerical results, the interference effect is uninfluenced by disorder in a large region. The reason is that the arising of MFs is protected by topology. Thus, the interference effect is always existing unless the MFs is destroyed by disorder. In supplement Fig. 2 we consider the case with the disorder strength is $V_{imp}=4\Delta$ which is equally distributed at range $[-V_{imp}/2,V_{imp}/2]$. Supplement Fig. 2(a) shows the contour plot of  total differential conductance from left lead to right lead with incident energy $E$ and chemical potential $\mu$. The information of interference effect can be well observed through the total conductance. It is the same as the main text that the total conductance can be described by two parts: $T_e$ contributed by electron transmission and $T_h$ contributed by CAR.  Supplement Fig. 2(b) gives the contour plot of electron transmission coefficient $T_{e}$, We can see that the ET process displays the interference effect well. While the contour plot of crossed Andreev reflection $T_h$  still hardly see the information of interference due to the self-hermitian of MFs.

\begin{figure}
\centering
\includegraphics[width=3.25in]{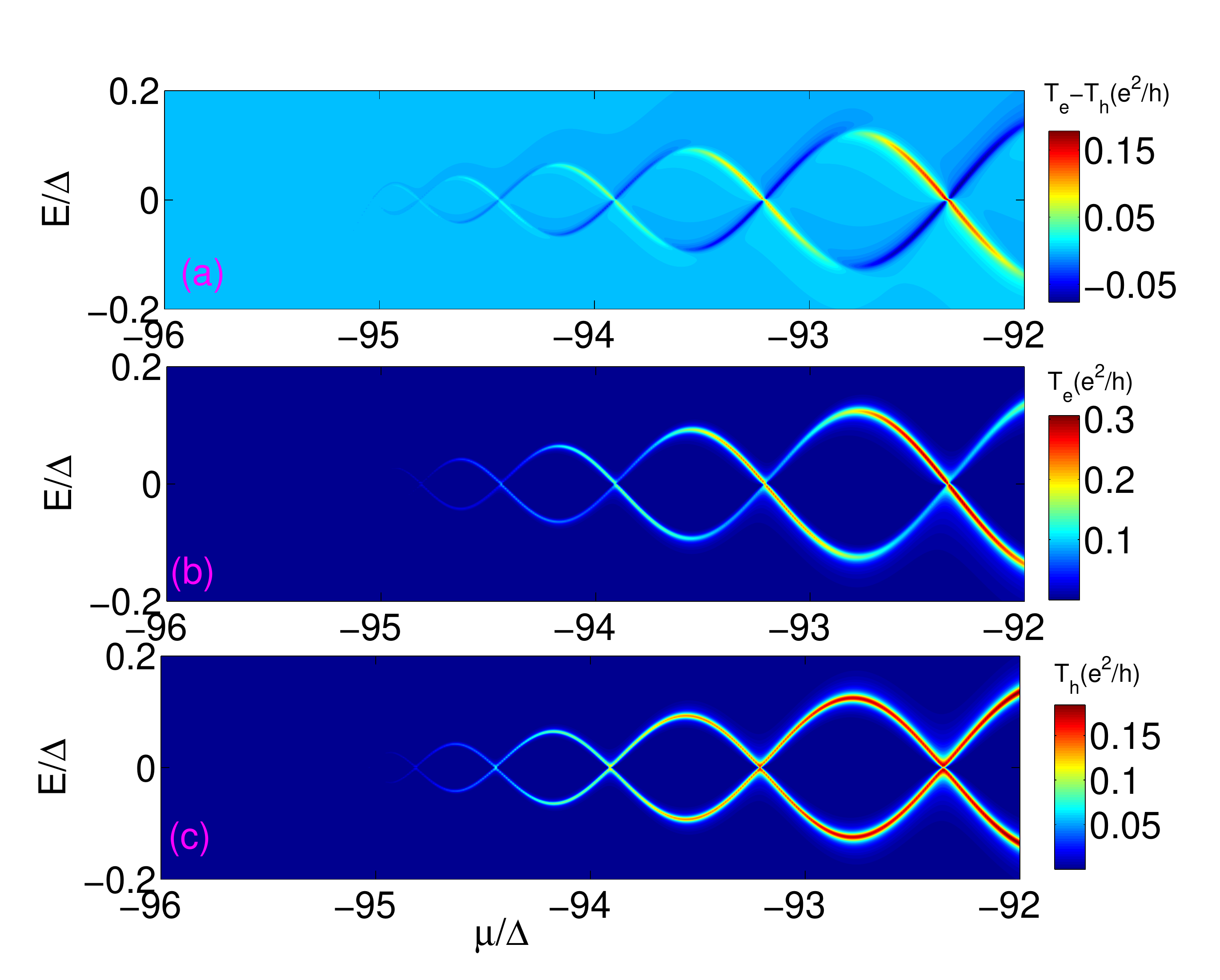}
\caption{ The case in quasi-2d situation, here we set $\mu\approx -4t$ which means it localized at the first topological region. (a) Contour plot of  total differential conductance from left lead to right lead with incident energy $E$ and chemical potential $\mu$.(b) Contour plot of electron transmission coefficient $T_{e}$. In this case the whole information is the same as the 1D case, we can watch the interference effect through the ET process well. (c) Contour plot crossed Andreev reflection $T_h$ as a function of  incident energy $E$  and chemical potential $\mu$.}
\end{figure}

\begin{figure}
\centering
\includegraphics[width=3.25in]{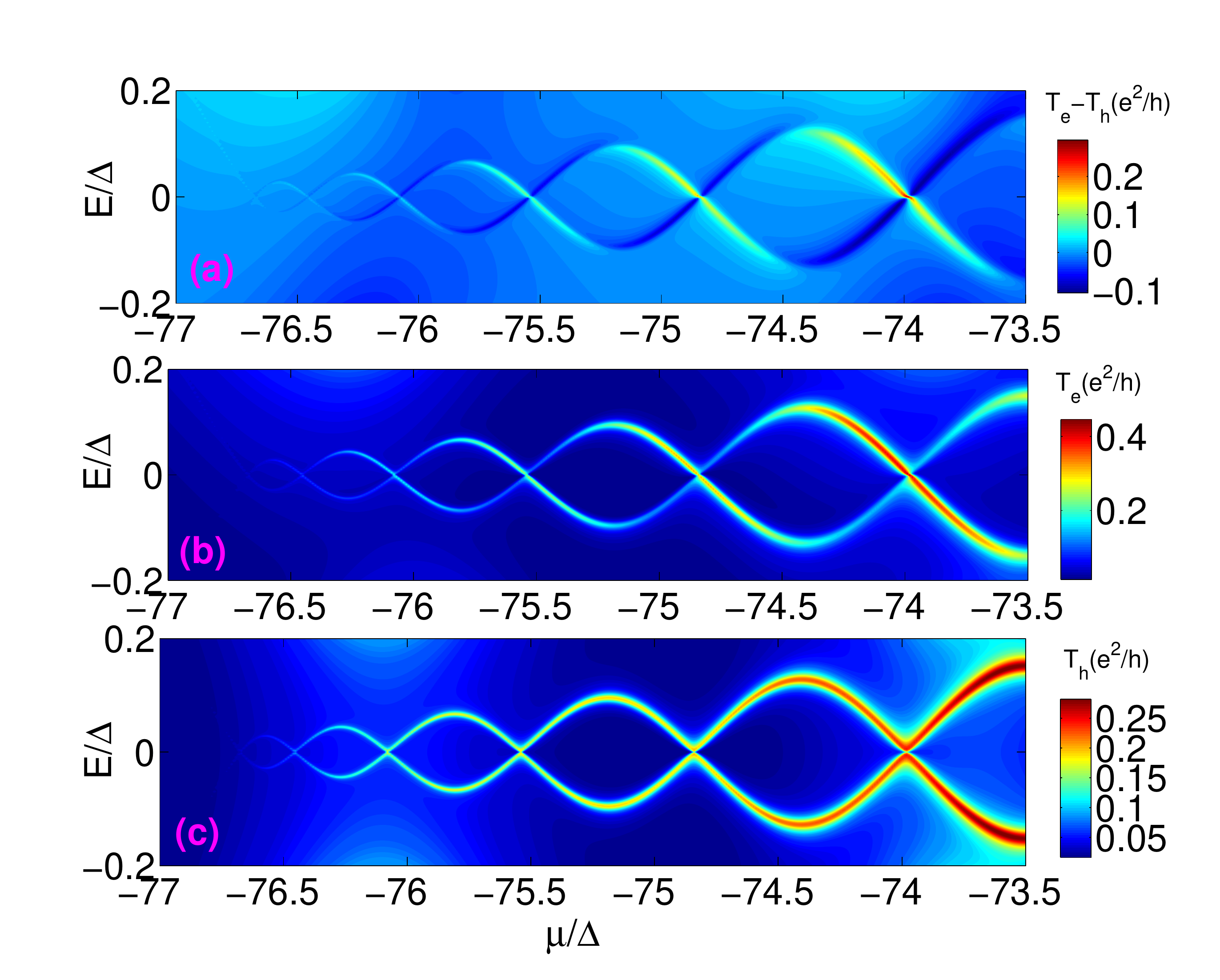}
\caption{The case in quasi-2d situation with the second topological region.(a) Contour plot of  total differential conductance from left lead to right lead with incident energy $E$ and chemical potential $\mu$. (b) Contour plot of electron transmission coefficient $T_{e}$. (c) Contour plot crossed Andreev reflection $T_h$ as a function of  incident energy $E$  and chemical potential $\mu$.   In this case  we can also observe the interference effect through the ET process well.} \label{2Dcurrent}
\end{figure}

\begin{figure}
\centering
\includegraphics[width=3.25in]{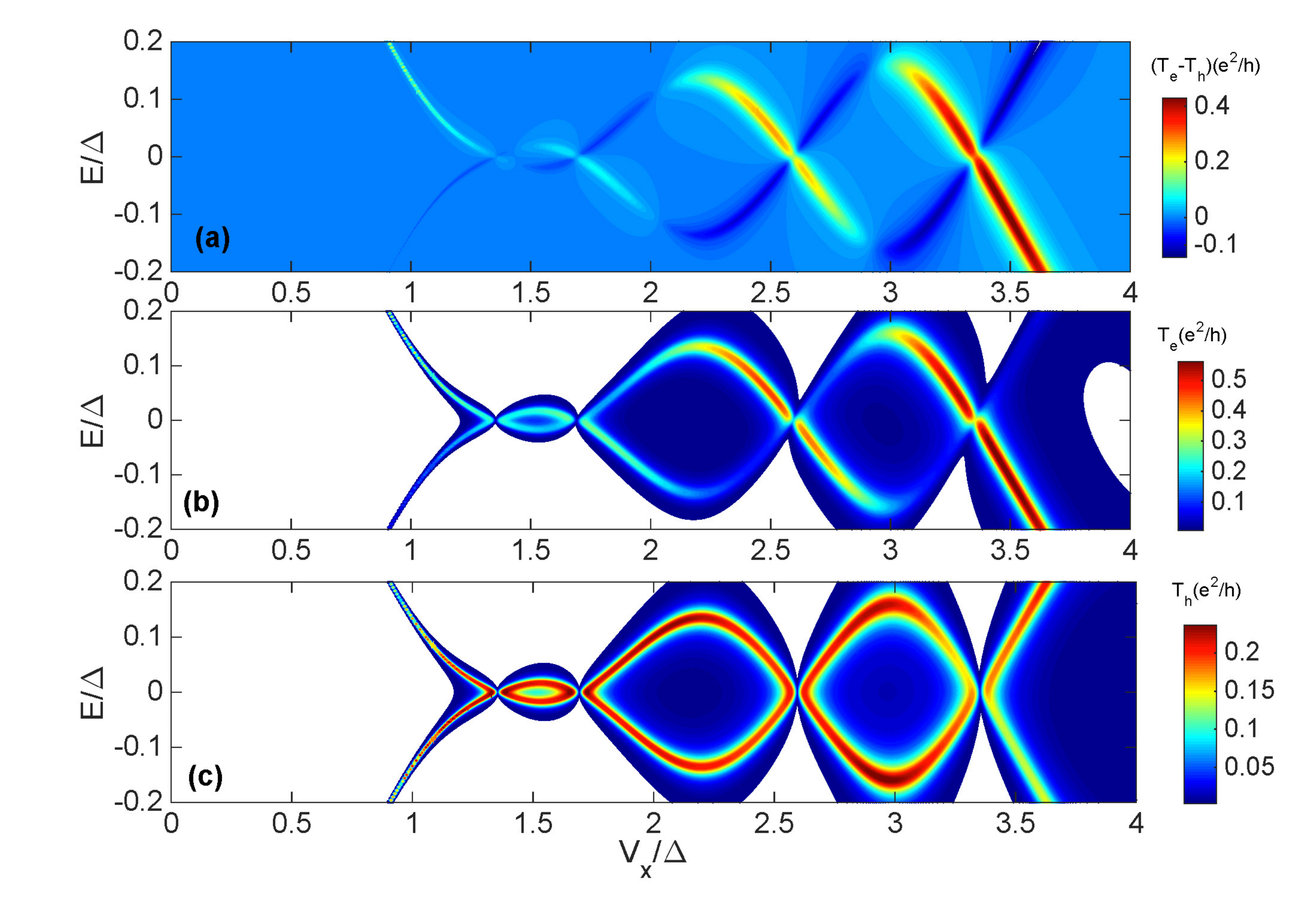}
\caption{The zeeman field can also adjust effective fermi wave vector well. Through adjusting the magnetic field we can also see the interference effect well. (a) Contour plot of  total differential conductance from left lead to right lead with incident energy $E$ and zeeman field $V_x$. (b) Contour plot of electron transmission coefficient $T_{e}$. (c) Contour plot crossed Andreev reflection $T_h$ as a function of  incident energy $E$  and zeeman field $V_x$. } \label{2Dcurrent}
\end{figure}

{\bf \emph{Results using a two dimensional model}} --- In the main text we use the strictly 1D wire to stress our main point. However, it is also quasi-2d or quasi-3d case in real system.
In such case, we can see several topological region as shown in our previous result. Then a question arises as to: the interference effect would be influenced when considering the quasi-2d or quasi-3d case. In supplement material Fig. 3 and Fig. 4 shows the results in higher dimensional case. Fig. 3 shows the results in first topological region while Fig. 4 shows the results in second topological region. In both case we can see that the ET process shows the information of interference effect well but the CAR process is insensitive to the interference effect.

{\bf \emph{Results using zeeman field.}} ---From the main text we have known that both the chemical potential and zeeman field can adjust the effective fermi wave vector well. Thus adjusting zeeman field can also change phase information well, then the interference effect can also be displayed by adjusting through a magnetic field along the wire. supplement Fig. 5 shows the results of the Contour plot of  $T_h$ and $T_e$. The interference effect is clear in this case.


\begin{thebibliography} {99}
\bibitem{note2} Considering that the two ends states lie at the ends of the wire should be two degenerate states:
 $\left(1 , 1 \right)^{T}$  and  $\left( i, -i \right)^{T} $; there will an $\pi$ phase shift comparing with the results in real system.
\end{thebibliography}
\end{document}